\documentclass[prc,preprint,showpacs,showkeys,nofootinbib]{revtex4}%
\usepackage{graphicx}
\usepackage{bm}
\usepackage{epsfig}
\usepackage{amsmath}
\usepackage{amsfonts}
\usepackage{amssymb}%
\begin{document}
\title{Probing nuclear skins and halos with elastic electron scattering}
\author{C.A. Bertulani}
\email{bertulani@physics.arizona.edu}
\affiliation{Department of Physics, University of Arizona, Tucson, AZ 85721, USA}
\date{\today}

\begin{abstract}
I investigate the elastic electron scattering off nuclei far from the
stability line. The effects of the neutron and proton skins and halos on the
differential cross sections are explored. Examples are given for the charge
distribution in Sn isotopes and its relation to the neutron skin. The neutron
halo in $^{11}$Li and the proton halo in $^{8}$B are also investigated.
Particular interest is paid to the inverse scattering problem and its
dependence on the experimental precision. These studies are of particular
interest for the upcoming electron ion colliders at the GSI and RIKEN facilities.

\end{abstract}
\maketitle

\section{Introduction}

The use of radioactive nuclear beams produced by fragmentation in high-energy
heavy-ion reactions lead to the discovery of halo nuclei, such as $^{11}$Li,
about 20 years ago \cite{Ta85}. Nowadays a huge number of $\beta$-unstable
nuclei far from stability are being studied thanks to further technical
improvements. Unstable nuclei far from stability are known to play an
important role in nucleosynthesis. Detailed studies of the structure and their
reactions will have unprecedented impact on astrophysics \cite{BHM02}.

The first experiments with unstable nuclear beams aimed at determining nuclear
sizes by measuring the interaction cross section in high energy collisions
\cite{Ta85}. Successive use of this technique has yielded nuclear size data
over a wide range of isotopes. Other techniques, e.g. isotope-shift
measurements, have allowed to extract the charge size. The growth of a neutron
skin with the neutron number in several isotopes have been deduced from
nuclear- and charge-size data \cite{OST01}.

The charge (nucleon) density distribution can also be determined by electron
(hadron) scattering experiments. Among hadron scattering, proton elastic
scattering at intermediate energies is a good tool to probe the nucleon
density distributions, due to its larger mean-free path in the nuclear medium.
But, undoubtedly, it is the electron scattering off nuclei that provides the
most direct information about charge distribution, which is closely related to
the spatial distribution of protons \cite{Hof61}.

A technical proposal for an electron-heavy-ion collider has been incorporated
in the GSI/Germany physics program \cite{Haik05}. A similar program exists for
the RIKEN/Japan facility using Self-Confining Radioactive
Ion Targets (SCRITs) \cite{Sud01}. In both cases the main purpose is to
study the structure of nuclei far from the stability line. The advantages of
using electrons in the investigation of the nuclear structure are mainly
related to the fact that the electron-nucleus interaction is relatively weak.
For this reason multiple scattering effects are usually small and the
scattering process is described in terms of perturbation theory. Since the
reaction mechanism in perturbation theory is well under control the connection
between the cross section and quantities such as charge distributions,
transition densities, response functions etc., is well understood \cite{Hof56}.

Theoretically, the shape of the density distribution includes detailed
information on the internal nuclear structure. In the independent particle
shell model the density distribution is the squared sum of the single-particle
wave functions. No measurement of either nucleon distribution or the charge
distribution for short-lived radioactive nuclei has been made so far.

Under the impulse approximation, or plane wave Born approximation, the charge
form factor can be determined from the differential cross section of elastic
electron scattering. Since the charge distribution, $\rho_{ch}(r)$, is
obtained from the charge form factor by a Fourier transformation, one can
experimentally determine $\rho_{ch}(r)$ by differential cross-section
measurements covering a wide range of momentum transfer $q$. This leads to
information on the size and diffuseness when the charge form factor is
measured at least up to the first maximum. To accomplish this with a reasonable
measuring time of one week, a luminosity larger than 10$^{26}$ cm$^{-2}%
$s$^{-1}$ is required, for example for the $^{132}$Sn isotope \cite{Haik05}.

On the theoretical side the difference between the proton and neutron
distributions can be obtained in the framework of Hartree-Fock (HF) method
(see for example \cite{HL98}) or Hartree-Fock-Bogoliubov (HFB) method (see for
example \cite{Miz00,Ant05}). As a rule of thumb, a theoretical calculation of
the nuclear density is considered good when it reproduces the data on elastic
electron scattering. But some details of the theoretical densities might not
be accessible in the experiments, due to poor resolution or limited
experimental reach of the momentum transfer $q$.
Recent works have also looked at electron scattering from halo nuclei, see e.g.
refs. \cite{GMG99,Ers05,KA06}.

In this work I study the general features of elastic electron
scattering off unstable nuclei. Here I focus on using the general
features of what is theoretically known about skins and halos to
look for their respective signatures in the data. Only a few
specific and representative nuclear density cases are used. The
paper is organized as follows. After a general introduction of
elastic electron scattering in section 2, I show in section 3 how
the telltales of neutron and proton skins and halos become visible
in the differential cross sections for elastic electron scattering.
Section 4 presents a study of the inverse scattering problem, which
poses a challenge for extracting the charge density profiles from
the experimental data. The conclusions are presented in section 5.

\section{Elastic Electron Scattering}

In the plane wave Born approximation (PWBA) the relation between the charge
density and the cross section is given by%
\begin{equation}
\left(  \frac{d\sigma}{d\Omega}\right)  _{\mathrm{PWBA}}=\frac{\sigma_{M}%
}{1+\left(  2E/M_{A}\right)  \sin^{2}\left(  \theta/2\right)  }\ |F_{ch}%
\left(  q\right)  |^{2}, \label{PWBA}%
\end{equation}
where $\sigma_{M}=(e^{4}/4E^{2})\cos^{2}\left(  \theta/2\right)
\sin^{-4}\left(  \theta/2\right)  $ is the Mott cross section, the term in the
denominator is a recoil correction, $E$ is the electron total energy, $M_{A}$
is the mass of the nucleus and $\theta$\ is the scattering angle.

The charge form factor $F_{ch}\left(  q\right)  $ for a spherical mass
distribution is given by%
\begin{equation}
F_{ch}\left(  q\right)  =4\pi\int_{0}^{\infty}dr\ r^{2}j_{0}\left(  qr\right)
\rho_{ch}\left(  r\right)  , \label{form}%
\end{equation}
where $q=2k\sin\left(  \theta/2\right)  $ is the momentum transfer, $\hbar k$
is the electron momentum, and $E=\sqrt{\hbar^{2}k^{2}c^{2}+m_{e}^{2}c^{4}}$.
The low momentum expansion of eq. \ref{form} yields the leading terms
\begin{equation}
F_{ch}\left(  q\right)  /Z=1-\frac{q^{2}}{6}\left\langle r_{ch}^{2}%
\right\rangle +\cdots. \label{formexp}%
\end{equation}
Thus, a measurement at low momentum transfer yields a direct assessment of the
mean square radius of the charge distribution, $\left\langle r_{ch}%
^{2}\right\rangle ^{1/2}$. However, as more details of the charge distribution
is probed more terms of this series are needed and, for a precise description
of it, the form factor dependence for large momenta $q$ is needed.

A theoretical calculation of the charge density entering eq. \ref{form} can be
obtained in many ways. Let $\rho_{p}\left(  \mathbf{r}\right)  $ and $\rho
_{n}\left(  \mathbf{r}\right)  $\ denote the point distributions of the
protons and the neutrons, respectively, as calculated, e.g. from
single-particle wavefunctions obtained from an average one-body potential
well, the latter in general being different for protons and neutrons. If
$f_{Ep}\left(  \mathbf{r}\right)  $ and $f_{En}\left(  \mathbf{r}\right)  $
are the spatial charge distributions of the proton and the neutron in the
non-relativistic limit, the charge distribution of the nucleus is given
by%
\begin{equation}
\rho_{ch}\left(  \mathbf{r}\right)  =\int\rho_{p}\left(  \mathbf{r}^{\prime
}\right)  \ f_{Ep}\left(  \mathbf{r-r}^{\prime}\right)  \ d^{3}r^{\prime}%
+\int\rho_{n}\left(  \mathbf{r}^{\prime}\right)  \ f_{En}\left(
\mathbf{r-r}^{\prime}\right)  \ d^{3}r^{\prime}. \label{rhoch}%
\end{equation}

The second term on the right-hand side of eq. \ref{rhoch} plays an important
role in the interpretation of the charge distribution of some nuclear
isotopes. For example, the half-density charge radius increases 2\% from
$^{40}$Ca to $^{48}$Ca, whereas the surface thickness decreases by 10\% with
the result that there is more charge in the surface region of $^{40}$Ca than
of $^{48}$Ca \cite{Fro68}. This also implies that the rms charge radius of
$^{48}$Ca is slightly smaller than that of $^{40}$Ca. The reason for this
anomaly is that the added $f_{7/2}$ neutrons contribute negatively to the
charge distribution in the surface and more than compensate for the increase
in the rms radius of the proton distribution.

For the proton the charge density $f_{Ep}\left(  r\right)  $ in eq.
\ref{rhoch} is taken as an exponential function, corresponding to a form
factor $\widetilde{f}_{Ep}\left(  q\right)  =(1+q^{2}%
/\Lambda^{2})^{-1}$ (see Appendix A).
For the neutron a good parametrization is $\widetilde{f}_{En}%
(q)=-\mu_{n}\tau \widetilde{f}_{Ep}(q)/(1+p\tau)$,
where $\mu_{n}$ is the neutron magnetic
dipole moment and $\tau=q^{2}/4m_{N}$. We will use $\Lambda^{2}=0.71 $
fm$^{-2}$ (corresponding to a proton rms radius of 0.87 fm) and $p=5.6$, which
Galster et al. \cite{Gal71} have shown to reproduce electron-nucleon
scattering data.

Eqs. \ref{PWBA}-\ref{rhoch} are based on the first Born
approximation. They give good results for light nuclei (e.g.
$^{12}$C) and high-energy electrons. For large-$Z$ nuclei the
agreement with experiment is only of a qualitative nature. The
effects of distortion of the electron waves have been studied by
many authors (see, e.g. ref. \cite{MF48,Yenn53,Cut67}). More
important than the change in the normalization of the cross section
is the displacement of the minima. It is well known that a
simple modification can be included in the PWBA equation reproducing
the shift of the minima to lower $q$'s. One replaces the momentum
transfer $q$ in the form factor of eq. \ref{PWBA} with the effective
momentum transfer $q_{eff}=q\left(  1+3Ze^{2}/2R_{ch}E\right) $,
where $E$ is the electron energy and $R_{ch}\simeq1.2\ A^{1/3}$ fm. This
is because a measurement at momentum transfer $q$ probes in fact \
$|F(q)|^{2}$ at $q=q_{eff}$ due to the attraction the electrons feel
by the positive charge of the nucleus. This expression for $q_{eff}$
assumes a homogeneous distribution of charge within a sphere of
radius $R_{ch}$.

A realistic description of elastic electron scattering cross
sections requires full solution of the Dirac equation. The Dirac
equation for elastic scattering from a charge distribution can be
found in standard textbooks, e.g. \cite{EG88}. Numerous DWBA codes
based on Dirac distorted waves have been developed and are public.
Since the inclusion of Coulomb distortion is straight-forward, we
will concentrate on the information which can be extracted from
elastic electron scattering, which is encoded in the form-factor of
eq. \ref{form}. Later we will study the problem of extracting the
full nuclear density from the form factor. For simplicity, we will
also neglect the effect of the charge distribution of the neutron
itself. This effect is particularly important for large
$q$-transfers. For heavy nuclei it can change $\left\vert
F_{ch}\left(  q\right)  \right\vert ^{2}$\ by 10-20\% in the q-range
of 1.5 - 2 fm$^{-1}$.

\section{Skins and Halos}

\subsection{Neutron Skins}

Appreciable differences between neutron and proton radii are
expected \cite{Dob96} to characterize nuclei at the border of
the stability line. The liquid drop formula expresses the binding
energy of a nucleus with $N$ neutrons and $Z$ protons as a sum of
bulk,
surface, symmetry and Coulomb energies ${E/A}=a_{V}A-a_{S}A^{2/3}%
-S(N-Z)^{2}/A-a_{C}Z^{2}/A^{1/3}\mp a_{p}A^{-1/2},$where $a_{V}$, $a_{S}$,
$a_{p}$, $S$ and $a_{C}$ are parameters fitted to experimental data of
binding energy of nuclei. This equation does not distinguish between surface
($S$) and volume ($V$) symmetry energies. As shown in ref. \cite{Da03}, this
can be achieved by partitioning the particle asymmetry as $N-Z=N_{S}%
-Z_{S}+N_{V}-Z_{V}$. The total symmetry energy $S$ \ then takes on
the form
$S=S_{V}(N_{V}-Z_{V})^{2}/A+S_{S}(N_{S}-Z_{S})^{2}/A^{2/3}.$
Minimizing under fixed $N-Z$ leads to an improved liquid drop
formula \cite{Da03} with the term $S(N-Z)^{2}/A$ replaced by
\begin{equation}
S_{V}{\frac{(N-Z)^{2}}{A[1+\left(  S_{V}/S_{S}\right)  A^{-1/3}]}}.
\end{equation}

The same approach also yields a relation between the neutron skin $
R_{np}=R_{n}-R_{p}$, and $S_{S}$, $S_{V}$, namely \cite{Da03}%
\begin{equation}
\frac{R_{n}-R_{p}}{R}=\frac{A}{6NZ}(N_{S}-Z_{S})=\frac{A}{6NZ}\frac
{N-Z-\left(  a_{C}/12S_{V}\right)  ZA^{2/3}}{1+\left(  S_{S}/S_{V}\right)
A^{1/3}}, \label{deltaD}%
\end{equation}
where $R=(R_{n}+R_{p})/2$, and $R_n$ ($R_p$) is the half-density
neutron (proton) distribution radius.

Here the Coulomb contribution is essential; e.g. for $N=Z$ \ the neutron skin
$R_{np}$ is negative due to the Coulomb repulsion of the protons. A wide
variation of values of $S_{V}$ and $S_{S}$ can be found in the literature.
These values have been obtained by comparing the above predictions for energy
and neutron skin to theoretical calculations of nuclear densities and
experimental data on other observables \cite{Da03,St05}.
The values of $a_{C}=0.69$ MeV and $S_{V}=28$ MeV,
which will be used here, are compatible with fits of the binding energy and symmetry energy of known
nuclei \cite{St05}. Using these values and $R=1.2A^{1/3}$ fm, one gets, for large
$A$ (with $A\gg1$, $NZ\simeq A^{2}/4$),
\begin{equation}
R_{np}=R_{n}-R_{p}\simeq0.8\left(  \frac{S_{V}}{S_{S}}\right)  \left(
\delta-2.05\times10^{-3}ZA^{-1/3}\right)  \text{ fm},\label{Pawelh}%
\end{equation}
where $\delta=\left(  N-Z\right)  /A$ is the asymmetry parameter. The ratio
$S_V/S_S$ varies within the range $2 \lesssim S_S/S_V\gtrsim 3$ by adjusting eq.
\ref{deltaD} to the dependencies of skin sizes on asymmetry and on the
neutron-proton separation-energy difference constraints  \cite{St05}.
If one assumes that the central densities for neutrons and protons are
roughly the same and that they are both described by a uniform
distribution with sharp-cutoff radii, $R_{n}$ and $R_{p}$, one finds
$R_{np}\simeq 0.8A^{1/3}\delta$ fm. This shows that the sharp sphere
model is too simple and that in this model the central densities for
proton and neutrons have to differ if eq. \ref{deltaD} is to remain valid.

On the experimental front, a study of antiprotonic atoms published
in reference \cite{Trz01} obtained the following fitted formula for
the neutron skin of stable nuclei in terms of the root mean square
(rms) radii of protons and neutrons
\begin{equation}
\Delta r_{np}=\left\langle r_{n}^{2}\right\rangle ^{1/2}-\left\langle
r_{p}^{2}\right\rangle ^{1/2}=\left(  -0.04\pm0.03\right)  +\left(
1.01\pm0.15\right)  \delta\ \text{fm.}\label{delta}%
\end{equation}
The relation of the mean square radii with the half-density radii
for a Fermi distribution is roughly given by
$\left\langle r_{n}^{2}\right\rangle =3R_{n}^{2}/5+7\pi^{2}a_{n}^{2}/5$, where
$a_{n}$ is the diffuseness parameter. For heavy nuclei, assuming $a_{n}%
=a_{p}\ll R_{n},R_{p}$, one gets approximately the same linear dependence
on the asymmetry parameter as in eq. \ref{Pawelh} (neglecting the small
contribution of the $A^{-1/3}$ term).

In this work we will use eq. \ref{delta} as the starting point for
accessing the dependence of electron scattering on the neutron skin
of heavy nuclei. Applying it to calcium isotopes as an example, one
obtains that the neutron skin varies from $-0.15$ fm for $^{35}$Ca
(proton-rich with negative neutron skin) to $0.25$ fm for $^{53}$Ca.
A negative neutron skin obviously means an excess of protons at the
surface. For unstable nuclei no study as the one leading to eq.
\ref{delta} is experimentally available. Nucleon knockout reactions
with secondary unstable nuclear beams, and other techniques, have
been used to determine the interaction radii of neutron rich nuclei
\cite{Ta85}. These radii are thought to represent the extension of
the neutron distribution. But very little is known about the
charge distribution in neutron-rich nuclei.
\begin{figure}[ptb]
\begin{center}
\includegraphics[
height=3.1185in,
width=3.7126in
]{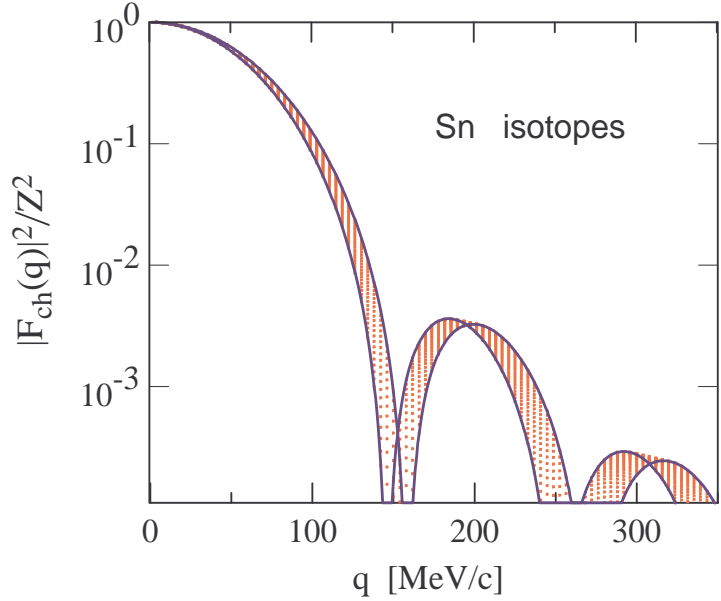}
\end{center}
\caption{The charge form factor squared for elastic electron
scattering off tin isotopes, as a function of the momentum transfer. The two
curves are for the extreme values of the asymmetry parameter $\delta=(N-Z)/A$, that is
$\delta=0$ ($N=Z=50$), and $\delta=4/9$ ($N=90$).
\ The curves form an envelope around other curves with intermediary values of
$\delta$.}%
\label{form2}%
\end{figure}

For heavy nuclei the charge and neutron distributions can be described by a
Fermi distribution. The diffuseness is usually much smaller than the
half-density radius, $a_{p,n}\ll R_{p,n}$. The neutron skin is then given by
$R_{np}=R_{n}-R_{p}\simeq\sqrt{5/3}\ \Delta r_{np}$. We assume further that
the nuclear charge radius is represented by $R_{p}=1.2A^{1/3}%
\ \mathrm{fm}-R_{np}/2$ with $\Delta r_{np}$ given by eq. \ref{delta}. Figure
\ref{form2} shows the charge form factor squared for
elastic electron scattering off tin isotopes, as a function of the momentum
transfer. The two solid curves are for the extreme values of the asymmetry
parameter $\delta=(N-Z)/A$,  that is
$\delta=0$ ($N=Z=50$), and $\delta=4/9$ ($N=90$).
The curves form an envelope around other curves
with intermediary values of $\delta$. The first and second minima of the form
factors occur at $q_{1}=4.49/R_{p}$ and $q_{2}=7.73/R_{p}$, respectively,
corresponding to the zeroes of the transcendental equation $\tan\left(
qR_{p}\right)  =qR_{p}$.

Using the approximations discussed in the above paragraph, i.e.
$R_{p}=1.2A^{1/3}\ \mathrm{fm}-R_{np}/2$ with $\Delta r_{np}$, $R_{np}= \Delta r_{np}$,
and the experimental value given in eq. \ref{delta}, we conclude that
the linear dependence of $R_{p}$ with the neutron skin (and with the asymmetry
parameter $\delta$), also implies a linear dependence of the position of the
minima,
\begin{equation}
q_{1}\simeq\frac{3.74}{A^{1/3}}\ \ \left[  1-0.535\frac{(N-Z)}{A^{4/3}%
}\right]  ^{-1}\ \mathrm{fm}^{-1},\ \ \ \ \ \ \ \ \mathrm{and}%
\ \ \ \ \ \ q_{2}=1.72\ q_{1}\ .
\end{equation}
For $^{100}$Sn the first minimum is expected to occur at $q_{1}=0.806$
$\mathrm{fm}^{-1}=159$ MeV/c, while for $^{132}$Sn it occurs at $q_{1}=0.754$
$\mathrm{fm}^{-1}=149$ MeV/c.

\begin{figure}[ptb]
\begin{center}
\includegraphics[
height=2.7735in,
width=3.4592in
]{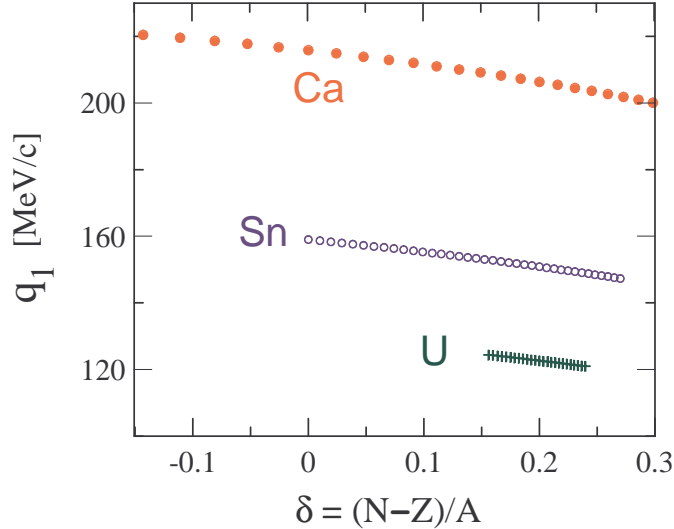}
\end{center}
\caption{Position of the first minimum $q_{1}$ for elastic electron scattering
off calcium, tin and uranium isotopes, as a function of the asymmetry
parameter $\delta$.}%
\label{q1min1}%
\end{figure}

Figure \ref{q1min1} shows the value of $q_{1}$ for calcium, tin and uranium
isotopes, as a function of the asymmetry parameter $\delta$. The variation of
$q_{1}$ with the neutron skin of neighboring isotopes, $\Delta q_{1}%
\simeq2/A^{8/3}$ fm$^{-1}$,  is too small to be measured accurately. The first
minimum, $q_{1}$, changes from $220$ MeV/c for $^{35}$Ca to $204$ MeV/c for
$^{53}$Ca, approximately 7\%, which is certainly within the experimental
resolution. Of course, sudden changes of the neutron skin with $\delta$\ might
happen due to shell closures, pairing,  and other microscopic effects.

To be more specific, let us assume that a reasonable goal is to
obtain accurate results for the charge radius $\left\langle
r_{p}^{2}\right\rangle ^{1/2},$ so that $\delta\left\langle
r_{p}^{2}\right\rangle ^{1/2}<0.05$ fm. This implies that the
measurement of $q_{1}$ has to be such that $\left( \Delta
q_{1}/q_{1}\right)  <q_{1}\left[  \text{fm}^{-1}\right]  \%$, with
$q_{1}$ in units of fm$^{-1}$ and the right-hand side of the
inequality yielding the percent value. For $^{53}$Ca, one has q$_{1}=$
$1.11$ fm$^{-1}$, meaning that the experimental resolution on the
value of $q_{1}$ has to be within 1\% if $\delta\left\langle
r_{p}^{2}\right\rangle ^{1/2}<0.05$ fm is a required precision. The
situation improves for heavier nuclei, as becomes evident from
figure \ref{q1min1}.

The neutron skin dependence is also seen in the height of the second
bump, after the first minimum. This bump occurs at
$q_{m}\simeq6/R_{p}$. For a Woods-saxon density this peak will be
reduced compared to the first maximum by a factor
$[g(q_{m}a)/12]^{2}$, where $g(q_{m}a)$ is a function of the
diffuseness parameter $a$. $g$ is closely given by the Fourier
transform of an Yukawa function, its value at $q_{m}a$ is of the
order of $1/2$ and its variation around $q_{m}$ is weak,
$g(qa)\sim1/\left(  1+q^{2}a^{2}\right)  $ (see Appendix A). Thus, the dependence of
the height of the second maximum upon the neutron skin is a less
appropriate tool than the location of the first minimum. Of course,
the ultimate test of a given theoretical model will be a good
reproduction of the measured data, below and beyond the first
minimum.

\subsection{Neutron halos}

\begin{figure}[ptb]
\begin{center}
\includegraphics[
height=2.8963in,
width=3.5708in
]{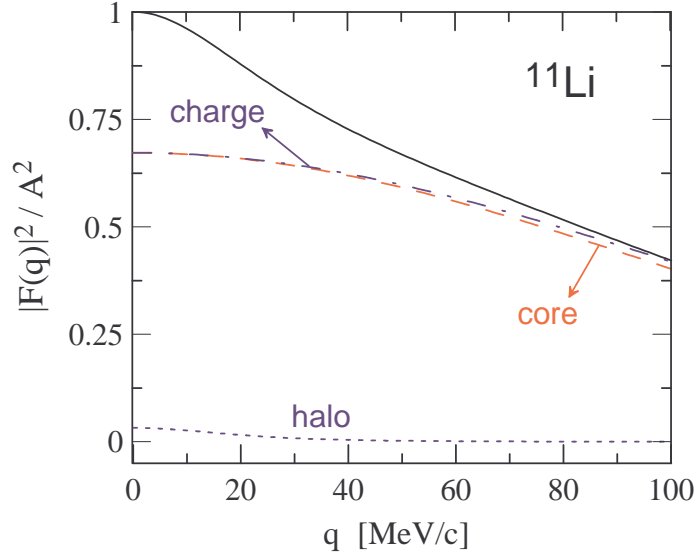}
\end{center}
\caption{Form factor, $|F(q)|^2/A^2$, for $^{11}$Li showing the individual
contributions of the core and of the halo nucleons. \ The dotted
(halo), dashed (core) and solid (total) curves are form factors for
the matter distribution. The dashed-dotted curve is the charge form factor,
$|F_{ch}(q)|^2/Z^2$.}%
\label{form3}%
\end{figure}

For light halo
nuclei composed of a core nucleus and an extended distribution of
halo nucleons, the nuclear matter form factor can be fitted with the
simple
expression%
\begin{equation}
F(q)/A=\left(  1-g\right)  \exp\left(  -q^{2}a_{1}^{2}/4\right)  +\frac
{g}{1+a_{2}^{2}q^{2}},\label{haloform}%
\end{equation}
where $g$ is the fraction of nucleons
in the halo. In this expression the first term \ follows from the
assumption that the core is described by a Gaussian and the second term
assumes that the halo nucleons are described by an Yukawa distribution
(see Appendix A). Taking $^{11}$Li as an example,
the following set of parameters can be used $g=0.18,$ $a_{1}=2.0$ fm
and $a_{2}=6.5$ fm. This means that $g=2/11$ nucleons are in the halo,  the
size of the core is roughly 2 fm, and the size of the halo is 6.5 fm.
Although only few nucleons are in the halo they
change dramatically the shape of the form factor, as
shown in figure \ref{form3}. The dashed-dotted curve is the squared charge form
factor. The dotted curve applies for the matter density of the halo
neutrons. When the core (dashed curve) and halo nucleon distributions are combined
the squared form factor for the total matter distribution (solid curve) clearly
displays the halo signature. Thus, even
when the individual contribution of the halo nucleons is small and
barely visible in a linear plot of the matter distribution, it is
very important for the form factor of the total matter distribution.
It is responsible for the narrow peak which develops at low momentum
transfers. This signature of the halo was indeed the motivation for
the early experiments with radioactive beams. The narrow peak was
observed in momentum distributions following knockout reactions
\cite{Ta85}.

Elastic electron scattering will not be sensitive to the narrow peak
of $\left\vert F(q)\right\vert ^{2}$ (matter distribution form
factor) at small momentum $q$, but to the the charge distribution
form factor, $\left\vert F_{ch}(q)\right\vert ^{2},$ which in the
case of $^{11}$Li will be similar to the dashed-dotted curve in
figure \ref{form3}. The determination of this form factor will tell
us if the core has been appreciably modified due to the presence of
the halo nucleons. It is worthwhile mentioning that many of what we
call core nuclei are also short-lived and very little is known about
their charge distribution.

In order to explain the spin, parities, separation energies and size
of exotic nuclei consistently a microscopic calculation is needed.
One possibility is to resort to a Hartree-Fock (HF) calculation.
Unfortunately, the HF theory cannot provide the predictions for the
separation energies within the required accuracy of hundred keV.
Here we use a simple and tractable HF\ method ~\cite{BBS89} to
generate synthetic data for the charge-distribution of $^{11}$Li.
Details of this method is described in ref. \cite{SB97}. Assuming
spherical symmetry, the equation for the Skyrme interaction can be
written as \
\begin{equation}
\left[  -\mathbf{\nabla}\frac{\hbar^{2}}{2m^{\ast}(r)}\mathbf{\nabla
}+V(\mathbf{r})\right]  \psi_{i}(\mathbf{r})=\epsilon_{i}\psi_{i}%
(\mathbf{r})\label{schro}%
\end{equation}
where $\,m^{\ast}(r)\,$ is the effective mass. The potential $\,V(\mathbf{r}%
)\,$ has a central, a spin-orbit, and a Coulomb term
\begin{equation}
V(\mathbf{r})=V_{\mathrm{central}}+V_{\mathrm{spin-orbit}}+V_{\mathrm{Coulomb}%
}.\label{vr}%
\end{equation}
The central potential is multiplied by a constant factor $\,f\,$ only for the
last neutron configuration:%
\begin{equation}
V_{\mathrm{central}}(r)=fV_{HF}(r),\left\{
\begin{array}
[c]{l}%
f\neq1\quad\mathrm{for\;last\;neutron\;configuration}\\
f=1\quad\mathrm{otherwise.}%
\end{array}
\right.  \label{vcentral}%
\end{equation}
Thus, the last neutron configuration (last orbits) is treated
differently from the other orbits in the HF potential in order to
reproduce the neutron separation energy of the neutron-rich nucleus.
The factor \thinspace$f$ \thinspace in Eq.~\ref{vcentral} is
arbitrary ($f=0.82$ for the last neutrons in $^{11}$Li
\cite{BBS89}). It roughly scales with the inverse of the fraction of
nucleons in the halo, and simulates a weaker potential at the halo
region. This model was successful to explain most features of the
light-neutron rich nuclei \cite{BBS89,Sa92}. It can also explain the
magnitude of the nuclear sizes.  In order to obtain the nuclear
sizes, the  rms radii of the occupied nucleon orbits are multiplied
by the shell model occupation probabilities, which are also obtained
in the calculations. The final radius is obtained by adding the core
radius. It is important to notice that the physics of $^{11}$Li is
not treated very well because of the pairing interaction. This is
needed to make $^{11}$Li bound while $^{10}$Li is unbound. In this
aspect, the model adopted here is only useful as a qualitative tool.

\begin{center}%
\begin{tabular}
[c]{cccc} \hline\hline & $j_{\mathrm{halo}}$ & $\sqrt{\langle
r^{2}\rangle}_{\mathrm{cal}}\;$ (fm) & $\sqrt{\langle
r^{2}\rangle}_{\exp}\;$ (fm)\\\hline $^{9}$Li (core) &  & 2.45 &
2.43 $\pm$ 0.07$^\dagger$\\\hline
$^{11}$Li & $1p_{1/2}$ & 5.36 & \\
& $2s_{1/2}$ & 7.61 & \\
SM &  & 3.26 & 3.62 $\pm$ 0.19$^\dagger$\\\hline\hline $^\dagger$
from ref. \cite{Ege02}.
\end{tabular}
\end{center}
{\small Table 1\ -\ Single particle properties of }$^{11}$Li. \
{\small The second column gives the spins of the most probable
occupied orbits. The third column is the result of HF calculations
for the rms radii associated with these orbits, and the last column
gives the rms radii of the matter distribution of these nuclei.}

\medskip

As the effective interaction, a parameter set of the density
dependent Skyrme force, so called BKN interaction \cite{BKN}, is
adopted. The parameter set of BKN interaction has the effective mass
m$^{\ast}$/m =1 and gives realistic single particle energies near
the Fermi surface in light nuclei. The original BKN force has no
spin-orbit interaction. In the present calculations, I introduce the
spin-orbit term in the interaction so that the single-particle
energy of the last neutron orbit becomes close to the experimental
separation energy. In this way, the asymptotic form of the
loosely-bound wave function becomes realistic in the neutron-rich
nucleus. The large r.m.s. radii of the valence neutron orbits is
attributed to the small separation energy. The calculated value is
enough to create the halo structure in the HF wave functions. In
${^{11}}$Li, the last occupied orbits is taken to be 1p$_{1/2}$ and
2s$_{1/2}$.

We took the cut-off radius of H-F calculation to be $R=40\,\mbox{fm}$ which is
necessary to include properly the loosely bound nature of the neutron wave
function. The separation energy for the valence nucleons is S$_{2n}%
(theo)=0.274$ MeV, which is to be compared with the experimental one
S$_{2n}(exp)=0.247 \pm 0.080$ MeV \cite{TUNL}. The column indicated by $\,j_{\mathrm{halo}}\,$ in
Table 1 displays the most probably occupied orbits. The final radius is
obtained by adding the core radius, and is given in the row indicated by SM.

The elastic form factor for the matter distribution obtained in the HF
calculations are very close to the ones calculated by the empirical formula
\ref{haloform}. In figure \ref{form3} we show the charge form factor,
$\left\vert F_{ch}(q)\right\vert ^{2},$ by a dashed-dotted line. There is very
little difference from the simpler empirical parametrization of eq.
\ref{haloform}.  The lack of minima, and of secondary peaks (as in eq.
\ref{haloform}), makes it difficult to extract from $\left\vert F_{ch}%
(q)\right\vert ^{2}$ more detailed information on the charge-density
profile. Parametrizations like eq. \ref{haloform} which apparently
lack of a sound physical basis are not particular to loosely bound
nuclear systems. For example, in the case of $^{6}$Li a good fit to
experimental data was obtained with \cite{Sue67}
\[
\left\vert F_{ch}(q)\right\vert ^{2}\propto\exp(-a^{2}q^{2})-Cq^2\exp(-b^{2}%
q^{2}),
\]
with $a=0.933$ fm, $b=1.3$ fm, and $C=0.205$. However, the data  \cite{Sue67}
cannot be
fitted by using a model in which the nucleons move in a single-particle potential.

\subsection{Proton Halos}

\begin{figure}[ptb]
\begin{center}
\includegraphics[
height=2.7786in,
width=3.3382in
]{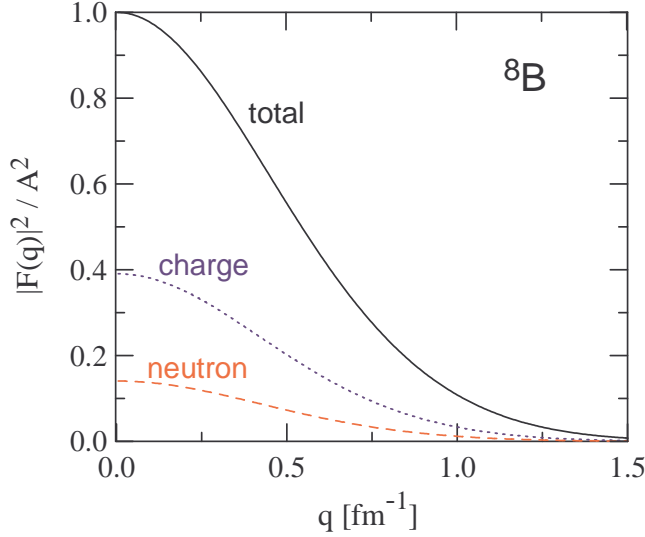}
\end{center}
\caption{Form factor squared for $^{8}$B showing the contribution of neutron,
$|F(q)|^2/N^2$,
(dashed) charge, $|F(q)|^2/Z^2$, (dotted) and total, $|F(q)|^2/A^2$, (solid) matter distribution of nucleons.}%
\label{formb81}%
\end{figure}

Here I will consider  $^{8}$B as a prototype of proton halo nucleus.
This nucleus is perhaps the most likely candidate for having a
proton halo structure, as its last proton has a binding energy of
only 137 keV. The charge density for this nucleus can be calculated
in the framework of the Skyrme HF model. We will use here the
results obtained in ref. \cite{Shyam}, where axially symmetric HF
equations were used with SLy4 \cite{Chab97} Skyrme interaction which
has been constructed by fitting the experimental data on radii and
binding energies of symmetric and neutron-rich nuclei. Pairing
correlations among nucleons have been treated within the BCS pairing
method. The form factor squared for the charge density in $^{8}$B is
shown in figure \ref{formb81}, normalized to mass number for matter
distribution, and to charge number for charge distribution.

The width of the charge form factor squared corresponding to the plot in
figure \ref{formb81} is $\Delta_{ch}=0.505$ fm$^{-1}=99.6$ MeV/c. The
corresponding values for the form factors for neutron and total matter distributions are,
respectively, $\Delta_{n}=0.512$ fm$^{-1}=101$ MeV/c and $\Delta_{tot}=0.545$
fm$^{-1}=108$ MeV/c. This amounts to an approximately 10\% difference between matter
and charge form factors in $^{8}$B. 

\begin{figure}[ptb]
\begin{center}
\includegraphics[
height=2.2684in,
width=2.2425in
]{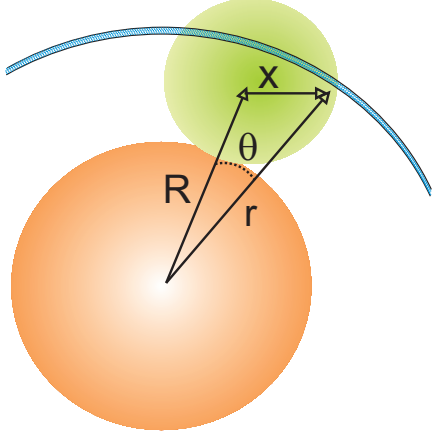}
\end{center}
\caption{A spherical shell with radius $r$ from the center of the
nucleus will have a contribution from the charge
inside a nucleon located at a distance $R$.}%
\label{nucleon}%
\end{figure}

The proton halo in $^{8}$B is mainly due to the unpaired proton in
the p$_{3/2}$ orbit. It is clear that for such a narrow halo the
size of the nucleon also matters. The relevance of the nucleon size
is shown in figure \ref{nucleon} where a slice of the nucleon is
included in a thin spherical shell of radius $r$ and thickness $dr$
from the center of the nucleus. If the position of the nucleon is
given by $R$, the part of the proton charge included in the
spherical shell is given by
\begin{equation}
d\rho_{ch}=2\pi r^{2}dr\int_{0}^{\pi}d\theta\ \rho_{p}\left(  \mathbf{x}%
\right)  \sin\theta,\label{drhon}%
\end{equation}
where $\rho_{p}\left(  \mathbf{x}\right)  $ is the charge distribution inside
a proton at a distance $\mathbf{x}$ from its center. The coordinates are shown
in figure \ref{nucleon}. They are related by $x^{2}=r^{2}+R^{2}-2rR\cos\theta
$. The contribution to the nuclear charge distribution from a single-proton in
this spherical shell is thus given by $\mathcal{N}_{p}\left(  R,r\right)
=d\rho_{ch}/4\pi r^{2}dr.$

Assuming that the charge distribution of the proton is described either by a
Gaussian or a Yukawa form, the integral in eq. \ref{drhon} can be performed
analytically, yielding%
\begin{equation}
\mathcal{N}_{p}^{(G)}\left(  R,r\right)  =\frac{1}{4\pi^{1/2}arR}\left\{
\begin{array}
[c]{c}%
\frac{1}{\pi}\left\{  \exp\left[  -\frac{\left(  R-r\right)  ^{2}}{a^{2}%
}\right]  -\exp\left[  -\frac{\left(  R+r\right)  ^{2}}{a^{2}}\right]
\right\}  ,\ \text{\ \ \ for a Gaussian dist.}\\
\frac{1}{2}\left\{  \exp\left[  -\frac{\left\vert R-r\right\vert }{a}\right]
-\exp\left[  -\frac{\left\vert R+r\right\vert }{a}\right]  \right\}
,\ \text{\ \ \ for a Yukawa dist.,}%
\end{array}
\right.  \label{npc}%
\end{equation}
where $a$ is the proton radius parameter.

The charge distribution at the surface of a heavy proton-rich nucleus,
$\delta\rho_{ch}\left(  r\right)  $, may be described as a pile-up of protons
forming a skin. Let $n_{i}$ be the number of protons in the skin and $R_{i}$
their distance to the center of the nucleus. One gets%
\begin{equation}
\delta\rho_{ch}\left(  r\right)  =\sum_{i}n_{i}\mathcal{N}_{p}\left(
r,R_{i}\right)  .\label{npcs}%
\end{equation}
Assuming $R_{i}$ to be constant, equal to the nuclear charge radius $R$, and
using eqs. \ref{npc} it is evident that while the density at the surface
increases, its size $R$ and width $a$, remain unaltered. \ The form factor
associated with this charge distribution is given by%
\begin{equation}
\delta F_{ch}\left(  q\right)  =\frac{4\pi}{q}\sum_{i}n_{i}\int_{0}^{\infty
}dr\ r\ \mathcal{N}_{p}\left(  r,R_{i}\right)  \ \sin\left(  qr\right)
=\exp\left(  -qa^{2}\right)  \sum_{i}n_{i}\frac{\sin\left(  qR_{i}\right)
}{qR_{i}},\label{formg}%
\end{equation}
where the last result is for the Gaussian distribution. An analytical
expression can also be obtained for the Yukawa distribution. For $R_{i}=R$,
expression \ref{formg} shows that the increase of density in the skin does not
change the shape of the form factor, or of the cross section, but just its
normalization. The decrease of the form factor with $q$ is determined by $a$ alone,
and not by $n_{i}$. \ If the charge of additional protons is distributed
homogeneously across the nucleus including the skin, the form
factor will not change appreciably, except for a small change in $R$.

\begin{figure}[ptb]
\begin{center}
\includegraphics[
height=2.8591in,
width=3.499in
]{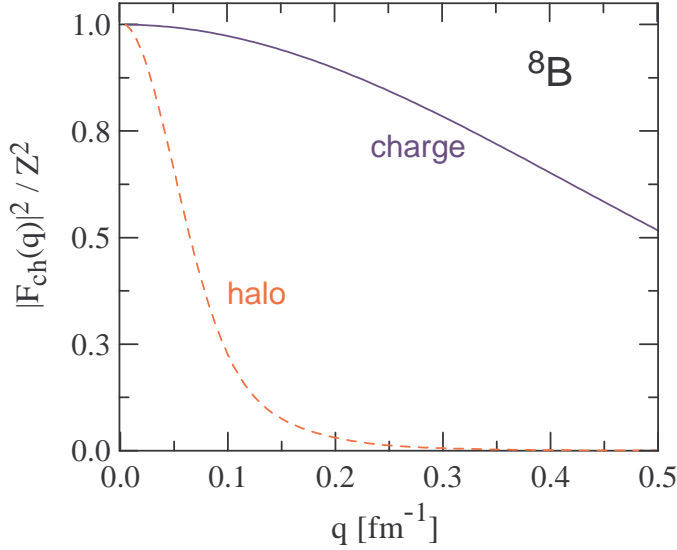}
\end{center}
\caption{The charge form factor squared of $^{8}$B (solid line),
$|F_{ch}^{total}(q)|^2/5^2$. Also shown is
the charge form factor squared due only to the valence halo proton,
$|F_{ch}^{halo}(q)|^2/1^2$.}%
\label{compf}%
\end{figure}

For a proton halo nucleus it is more appropriate to replace $\sum_{i}%
n_{i}\rightarrow4\pi\int dR\ R^{2}\ \mathcal{N}_{p}\left(  r,R\right)
\delta\rho_{ch}\left(  R\right)  $, where $\delta\rho_{ch}\left(  R\right)  $
is the density change created by the extended wavefunction of the halo
protons. We then recast eq. \ref{formg} in the form%
\begin{equation}
\delta F_{ch}\left(  q\right)  =\frac{4\pi}{q}\exp\left(  -qa^{2}\right)
\int_{0}^{\infty}dR\ R\ \delta\rho_{ch}\left(  R\right)  \ \sin\left(
qR\right)  .\label{formh}%
\end{equation}
The shape of the form factor is here dependent not only on the
proton size but also on the details of the halo density
distribution. For $^{8}$B, the halo size is determined by the
valence proton in a p$_{3/2}$ orbit. The density
$\delta\rho_{ch}\left(  R\right)  $\ due this proton can be
calculated with a Woods-Saxon model. Using the same potential
parameters as in ref. \cite{NBC06} we show in fig. \ref{compf} the
form factor $\left\vert \delta F_{ch}\left( q\right)  \right\vert
^{2}$ compared to the charge form factor $\left\vert F_{ch}\left(
q\right)  \right\vert ^{2}$ of figure \ref{formb81}. It is evident
that the halo contributes to a narrow form factor. However, in
contrast to the neutron halo case shown in figure \ref{form3}, the
charge form factor of $^{8}$B does not show a pronounced influence
of the halo charge distribution. 

The low energy expansion of the form factor, eq. \ref{formexp}, allows the
extraction of the rms radius of the charge distribution from%
\begin{equation}
\left\langle r_{ch}^{2}\right\rangle =-6\left[  \frac{dF_{ch}}{d\left(
q^{2}\right)  }\right]  _{q^{2}=0}. \label{derivf}%
\end{equation}

Applying this relationship to the charge form factor used in figure
\ref{compf} we get $\left\langle r_{ch}^{2}\right\rangle
^{1/2}=2.82$ fm which is close to the experimental value
$\left\langle r_{ch}^{2}\right\rangle _{\exp}^{1/2}=2.82\pm0.06$ fm
\cite{Blan97}.
The shape of the charge form factor can also be described by a
Gaussian distribution with radius parameter $a=2.30$ fm. In contrast
to the case of $^{11}$Li seen in figure \ref{form3}, the proton halo
in $^{8}$B does not seem to build up a two-Gaussian shaped form
factor. This observation also seems to be in contradiction with the
momentum distributions of $^{7}$Be fragments in knockout reactions
using $^{8}$B projectiles in high energy collisions \cite{Schw95}.
Electron scattering experiments will help to further elucidate this
property of proton halos.

\subsection{Comparison with inelastic scattering}

Loosely bound nuclei easily undergo breakup under any excitation. Such
inelastic processes will always accompany elastic scattering. We follow
the equations obtained in ref. \cite{Ber06} to make a crude estimate of the
effects of inelastic electron scattering off loosely bound nuclei.

The inelastic cross section for the
electro-excitation of loosely bound nuclei is given by \cite{Ber06}
\begin{equation}
\frac{d\sigma_{\rm inel}}{d\Omega dE_{x}}=\frac{48\sqrt{2}}{\pi}\frac{e^{2}\left[
e_{\mathrm{eff}}^{(1)}\right]  ^{2}p^{2}}{\hbar^{2}\mu c^{2}}\frac{1}{q^{2}%
}\frac{\sqrt{S_{n}}\left(  E_{x}-S_{n}\right)  ^{3/2}}{E_{x}^{4}%
},\label{sigeapp}%
\end{equation}
where $E_x$ is the excitation energy, $\Omega$ denotes the scattering solid angle,
$e$ is the electron charge, $e^{(1)}_{\rm eff}$ is the electric
effective charge for the nuclear response (I will assume for simplicity that
$e^{(1)}_{\rm eff}$=$Ze$), and $S_n$ is the nuclear breakup energy. $p$ is the
electron momentum, $q$ is the momentum transfer, and $\mu$ is the reduced mass,
where it is assumed that the nucleus is composed of two clusters.

We now integrate this equation over the excitation energy $E_x$, and obtain
\begin{equation}
\frac{d\sigma_{\rm inel}}{d\Omega }=\frac{3\sqrt{2}}{4}\frac{Z^2e^{4}}
{S_n\mu c^{2}}\frac{1}{\sin^2(\theta/2)%
}.\label{sigeapp2}%
\end{equation}
This equation is valid for large electron energies, so that $E\gg
m_ec^2$, and for small scattering angles, for which the
approximation $\hbar q\approx 2p \sin(\theta/2)$ can be used. In
fact, the minimum momentum transfer for an excitation energy $E_{x}$
is given by $q_{\min}=\Delta k\cong E_{x}/\hbar c$. Thus, the
inelastic  cross section tends to a maximum value at very small
scattering angles (in contrast, the elastic scattering cross section
increases indefinitely at small angles). Equation \ref{sigeapp2}
gives the behavior of the inelastic cross section when it starts to
decrease with angle, shortly after the maximum at zero degrees. A
characteristic feature emerging from eq. \ref{sigeapp2} is that the
inelastic cross section is proportional to the inverse of the
separation energy $S_n$.

From equation \ref{PWBA}, neglecting nuclear recoil, one gets for
the elastic scattering cross section $ d\sigma_{\rm el}/d\Omega
=Z^2e^4/4E^2\sin^4(\theta/2)$. The ratio between elastic and
inelastic cross sections for small scattering angles is thus given
by
\begin{equation}
\frac{d\sigma_{\rm el}}{d\sigma_{\rm inel}}=\frac{1}{3\sqrt{2}}
\frac{S_n\mu c^2}{E^2}
\frac{1}{\sin^2(\theta/2)%
}.\label{sigeapp4}%
\end{equation}
This shows that the elastic scattering dominates over inelastic
scattering for $\theta \ll \theta_{\max}\approx \sqrt{S_n \mu
c^2}/E$.  Adopting typical values, i.e. $E=100$ MeV, $S_n=1$ MeV and
$\mu c^2=10^3$ MeV, one gets $\theta_{\max}\approx 1/3$ radians. For
$E=1$ GeV this value reduces to $\theta_{\max}\approx 1/30$ radians.
These are kinematical constraints which have to be taken into
account in future electron scattering experiments.

\section{Inverse scattering problem}

In PWBA the inverse scattering problem can be easily solved. It is possible to
extract the form factor from the cross section and then, with an inversion of
the Fourier transform, to get the charge density distribution%
\begin{equation}
\rho_{ch}\left(  r\right)  =\frac{1}{2\pi^{2}}\int_{0}^{\infty}dq\ q^{2}%
j_{0}\left(  qr\right)  F_{ch}\left(  q\right)  .\label{densinv}%
\end{equation}
As we discussed in previous sections, the PWBA approximation can be justified
only for light nuclei (e.g. $^{12}$C) in the region far from the diffraction
zeros. For higher $Z$ values the agreement with experiment is only of a
qualitative nature.

It is very common in the literature to use a theoretical model for
$\rho _{ch}\left(  r\right)  $, e.g. the HF calculations discussed
in the previous sections and compare the calculated $F(q)$ with
experimental data. When the fit is \textquotedblleft
reasonable\textquotedblright\ (usually guided by the eye) the model
is considered a good one. However, whereas the theoretical
$\rho_{ch}\left(  r\right)  $ can contain useful information about
the central part of the density (e.g. bubble-like nuclei, with a
depressed central density), an excellent fit to the available
experimental data does not necessarily mean that the data is
sensitive to those details. The obvious reason is that short
distances are probed by larger values of $q$. Experimental data from
electron-ion colliders will suffer from limited accuracy at large
values of $q$, possibly beyond $q=1$ fm$^{-1}$. \ Thus it is useful
to identify what are the conditions for reproducing the nuclear
density within a theory independent fit.

In order to obtain an unbiased \textquotedblleft
experimental\textquotedblright\ $\rho_{ch}\left(  r\right)  $ one usually
assumes that the density is expanded as
\begin{equation}
\rho_{ch}\left(  r\right)  =\sum_{n=1}^{\infty}a_{n}f_{n}\left(  r\right)
,\label{dens}%
\end{equation}
where the basis functions $f_{n}(r)$ are drawn from any convenient complete
set and the expansion coefficients $a_{n}$ are adjusted to reproduce the
differential elastic cross section. The corresponding Fourier transform then
takes the form%
\begin{equation}
\widetilde{\rho}(q)\equiv F_{ch}\left(  q\right)  =\sum_{n=1}^{\infty}%
a_{n}\widetilde{f_{n}}\left(  q\right)  ,\label{rhoq}%
\end{equation}
where%
\begin{equation}
\widetilde{f_{n}}\left(  q\right)  =4\pi\int_{0}^{\infty}dr\ r^{2}j_{0}\left(
qr\right)  f_{n}\left(  r\right)  .\label{fnq}%
\end{equation}
represents basis functions in momentum space.

Evidently the sum in eq. \ref{dens} has to be truncated and this produces an
error in the determination of the charge density distribution. Another problem
is that, as shown by eq. \ref{densinv}, the solution of the inverse scattering
problem requires an accurate determination of the cross section up to large
momentum transfers. Electron scattering experiments in electron-ion colliders
will be performed within a limited range of $q$ and this will produce an
uncertainty in the determination of the charge density distribution. As we
have discussed in previous sections, the measurements have to encompass the
first few minima of the cross section for heavy nuclei in order that the
density profile can be mapped.

We consider two bases that have been found useful \cite{FN73} in the analysis
of electron or proton scattering data. The present discussion is limited to
spherical nuclei, but generalizations to deformed nuclei can be done. The
Fourier-Bessel (FB) expansion (i.e. with $f_{n}$ taken as spherical Bessel
functions) is useful because of the orthogonality relation between spherical
Bessel functions%
\begin{equation}
\int_{0}^{R_{\max}}dr\ r^{2}j_{l}\left(  q_{n}r\right)  j_{l}\left(
q_{m}r\right)  =\frac{1}{2}R_{c}^{3}\left[  j_{l+1}\left(  q_{n}R_{\max
}\right)  \right]  ^{2}\delta_{nm},\label{orthob}%
\end{equation}
where the $q_{n}$ are defined such as%
\begin{equation}
j_{l}\left(  q_{n}R_{\max}\right)  =0.\label{nodej}%
\end{equation}
The FB basis implies that the charge density $\rho_{ch}(r)$ should be zero for
values of $r$ larger than $R_{\max}$. For example, the basis can be defined as
follows
\begin{equation}
f_{n}\left(  r\right)  =j_{0}\left(  q_{n}r\right)  \Theta\left(  R_{\max
}-r\right)  \text{, \ \ \ \ \ \ \ \ \ \ \ \ \ }\widetilde{f_{n}}\left(
q\right)  =\frac{4\pi\left(  -1\right)  ^{n}R_{\max}}{q^{2}-q_{n}^{2}}%
j_{0}\left(  qR_{\max}\right)  \text{, \ }\label{FB}%
\end{equation}
where $\Theta$ is the step function, \ $R_{\max}$ is the expansion radius and
$q_{n}=n\pi/R_{\max}$.

In principle it is possible to obtain the $a_{n}$ coefficients measuring
directly the cross section at the $q_{n}$ momentum transfer. If the form
factor (\ref{form}) is known at $q_{n}$, the coefficients $a_{n}$ can be
obtained inserting (\ref{orthob}) and (\ref{FB}) in the definition
(\ref{form}) of the form factor, leading to%
\begin{equation}
a_{n}=\frac{F_{ch}\left(  q_{n}\right)  }{2\pi R_{\max}^{3}\left[
j_{1}\left(  q_{n}R_{\max}\right)  \right]  ^{2}}. \label{anFB}%
\end{equation}

In general the cross sections \ are measured at $q$ values different from
$q_{n}$. Using the expansion (\ref{FB}) of the charge density one finds for
the form factor the relation%
\begin{equation}
F_{ch}\left(  q\right)  =\frac{4\pi}{q}\sum_{n}a_{n}\frac{\left(  -1\right)
^{n}}{q^{2}-q_{n}^{2}}\sin\left(  qR_{\max}\right)  .\label{formFB}%
\end{equation}
By fitting the experimental $F_{ch}(q)$ one obtains the $a_{n}$ parameters and
reconstruct the nuclear charge density. Not all $a_{n}$'s are needed. Since
the integral of the density, or $F\left(  q=0\right)  $, is fixed to the
charge number there is one less degree of freedom. Also, densities tend to
zero at large $r$. Thus another condition can be used, e.g. that the
derivative of the density is zero at $R_{\max}$. \ Thus, when we talk about
$n$ expansion coefficients one means in fact that only $n-2$ coefficients need
to be used in eq. \ref{formFB}.\ For experiments performed up to $q_{\max}$
the number of expansion coefficients needed for the fit is determined by
$n_{\max}\simeq q_{\max}R_{\max}/\pi$.

A disadvantage of the FB expansion is that a relatively large number of terms
is often needed to accurately represent a typical confined density, e.g. for
light nuclei. One can use other expansion functions which are invoke less
number of expansion parameters, e.g. the Laguerre-Gauss (LG) expansion,
\[
f_{n}(r)=e^{-\alpha^{2}}L_{n}^{1/2}\left(  2\alpha^{2}\right)
,\ \ \ \ \ \ \ \ \ \ \ \ \ \ \ \ \ \text{and}\ \ \ \ \widetilde{f_{n}}%
(q)=4\pi^{3/2}\beta^{3}\left(  -1\right)  ^{n}e^{-\gamma^{2}}L_{n}%
^{1/2}\left(  2\gamma^{2}\right)  ,
\]
where $\alpha=r/\beta$, $\gamma=q\beta/2$, and $L_{n}^{p}$ is the generalized
Laguerre polynomial. Another possibility is to use an expansion on Hermite (H)
polynomials. In both cases, the number of terms needed to provide a reasonable
approximation to the density can be minimized by choosing $\beta$ in
accordance with the natural radial scale. For light nuclei $\beta=1-2$ fm can
be chosen, consistent with the parametrization of their densities. Then the
magnitude of $a_{n}$ decreases rapidly with $n$, but the quality of the fit
and the shape of the density are actually independent of $\beta$ over a wide
range.\ As shown by application to a few cases, the main information on skin
and halo sizes can be obtained using the FB expansion without problems.

\begin{figure}[ptb]
\begin{center}
\includegraphics[
height=2.9473in,
width=3.5509in
]{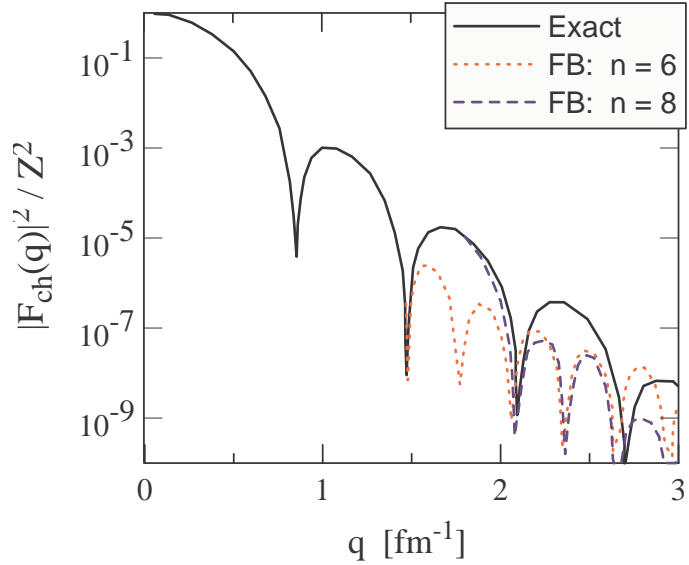}
\end{center}
\caption{Form factor squared for a Fermi function charge distribution with
half-density radius $R=5.2$ fm and diffuseness $a=1.2$ fm (solid line). The
dotted line is obtained with the FB expansion, eq. \ref{formFB}, up to $n=6$.
The dashed curve uses up to $n=8$.}%
\label{fb1}%
\end{figure}

Figure \ref{fb1} shows the charge form factor squared for a Fermi function
distribution with half-density radius $R=5.2$ fm and diffuseness $a=1.2$ fm
(solid line). The dotted line is obtained with the FB expansion, eq.
\ref{formFB}, up to $n=6$. The dashed curve uses up to $n=8$. One clearly sees
that the latter improves the fit to the form factor up to the third minimum by
increasing $n=6$ to $n=8$. We use $R_{\max}=15$ fm, so
that adding the $n=8$ term improves the fit of the distribution including the
peak at $q\simeq1.75$ fm$^{-1}$, as seen in the figure. 

For real data, the expansion coefficients $a_{n}$ are obtained by minimizing%
\[
\chi^{2}=\sum_{i}\left(  \frac{y_{i}-y(q_{i},a_{n})}{\sigma_{i}}\right)
^{2},
\]
where $y(q_{i},a_{n})$ is the fitted value of the cross section (form factor)
with a set of coefficients $a_{n}$ and $y_{i}$ are the experimental data at
momentum $q_{i}$ with uncertainty $\sigma_{i}$.

To check the limitations of this procedure, we generate a set of pseudodata
for $^{8}$B. The calculated charge form factor of figures \ref{formb81} and
\ref{compf} is used to generate 40 data points equally spaced by $\Delta
q=0.02$ fm$^{-1}$. \ These data were given an uncertainty $\sigma_{i}$
linearly increasing with $q_{i},$ from 1\% for $q_{i}=0.02$ fm$^{-1}$ to 20\%
for $q_{i}=0.82$ fm$^{-1}$.

\begin{figure}[ptb]
\begin{center}
\includegraphics[
height=2.4578in,
width=3.9859in
]{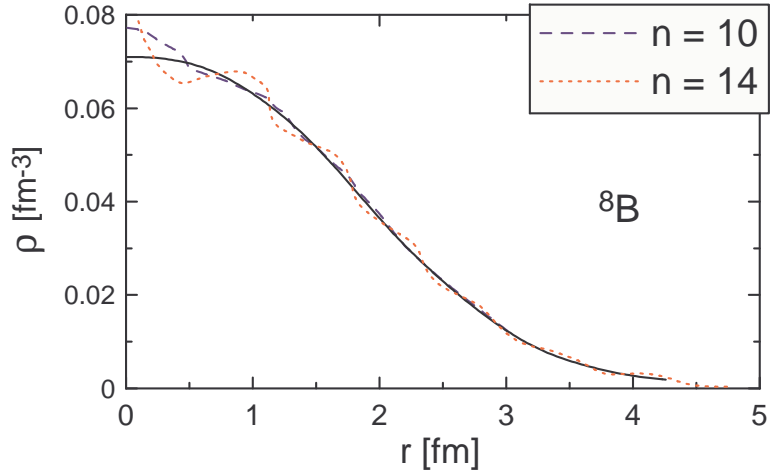}
\end{center}
\caption{ Hartree-Fock charge density of $^{8}$B (solid line). The dashed and
dotted lines are the solution of the inverse scattering problem using a
Fourier-Bessel expansion with $n=10$ and $n=14$ terms, respectively.}%
\label{fbb8}%
\end{figure}

The best fit, with $R_{\max}=10$ fm, is obtained with $n=10$ expansion
coefficients. Increasing the number of coefficients does not improve the
quality of the fit, as is shown in figure \ref{fbb8} for $n=14$. It only
produces more oscillations of the density. The reason is that terms with
larger $n$'s are only needed to reproduce the data at larger values of
momentum transfer, as shown in fig. \ref{fb1}. The fit to the data for
$q<q_{\max}$ is not affected but the presence of these new terms introduces
oscillations in the charge distribution. A possible fix to this problem is to
include pseudodata in addition to experimental data. This method is well known
in the literature \cite{FN73}. The pseudodata are used to enforce constraints
and to estimate the incompleteness error associated with the limitation of
experimental data to a finite range of momentum transfer.

\section{Conclusions}

In this work I have studied the electron scattering off light unstable nuclei.
This work is complementary to previous works in this area (see e.g. ref.
\cite{GMG99,Ant05,Ers05,KA06}). Particular attention was given to the effect of the neutron
(proton) skin on the scattering form factors. It was shown that the position
of the first minimum is arguably the best signature to look for noticeable
changes in the charge radius size.

The evidence of a proton halo is not so clear as in the case of other probes.
For example, it is well known that $^{7}$Be fragments arising from the proton
knockout of $^{8}$B projectiles display a distinctively narrow momentum
distribution characteristic of a long tail of the valence proton in $^{8}$B
\cite{Schw95}. This feature is a consequence of the peripheral character of
the knockout process, which is ideal to probe the tail of the bound states.
Nonetheless, due to the Coulomb barrier the bulk of the charge in the nucleus
is confined close to its center. Electron scattering is sensitive to this bulk
charge and therefore does not display such a strong halo signature.

The minimum information obtained with electron scattering in
electron-ion colliders will be the rms charge radius, $\left\langle
r_{ch}^{2}\right\rangle ^{1/2}$. This information \textquotedblleft
per se\textquotedblright\ is very valuable. It is sensitive to the
skin size in a heavy nucleus. But it also depends on the accuracy
with which this quantity can be measured. As we have shown in the
previous sections, it will be necessary to go beyond the first
minimum to extract information about the central value of the charge
density.

Accurate measurements at large momentum transfers are crucial if one
wants to describe the matter distribution with confidence and have a
good comparison with predictions of different theoretical models. I
have shown with a few basic examples using the Fourier-Bessel
expansion method that, whereas the matter distribution within the
halo is well probed by measurements at small momentum transfers, the
details of the central distribution requires measurements at large
$q$'s where inelastic processes may play an important role. This
again imposes constraints on the information that can be extracted
from elastic electron scattering off halo nuclei.

\section{Appendix 1 - Analytical form factors}

Here I summarize the analytical expressions for the charge form factors
which are used in this work.
For the nuclear charge density
given by a hard sphere (uniform distribution with sharp cutoff at $R$)
\begin{equation}
\rho\left(  r\right)  =\left\{
\begin{array}
[c]{c}%
\rho_{0}\text{ \ \ \ \ \ for \ \ \ }r\leq R\\
0\text{, \ \ \ \ \ \ \ otherwise,}%
\end{array}
\right.  \label{hards}%
\end{equation}
the form factor is%
\begin{equation}
F(q)=\frac{4\pi\rho_{0}}{q^{3}}\left[  \sin\left(  qR\right)  -qR%
\cos\left(  qR\right)  \right]  . \label{hardsf}%
\end{equation}

For an Yukawa distribution, $\rho\left(  r\right)  =\rho
_{0}\ e^{-r/a}(a/r)$, one gets
\[
F(q)=\frac{4\pi\rho_{0}a^{3}}{1+a^{2}q^{2}}.
\]

The symmetrized Fermi distribution
\cite{GLM91}%
\[
\rho(r)=\rho_{0}\frac{\cosh(R/a)}{\cosh(R/a)+\cosh(r/a)}
\]
leads to the form factor \cite{Co91}:%
\[
F(q)=-\frac{4\pi^{2}\rho_{0}a}{q}\frac{\cosh(R/a)}{\sinh(R/a)}\left[
\frac{R\cos(qR)}{\sinh\left(  \pi qa\right)  }-\frac{\pi d\sin
(qR)\cosh(\pi qa)}{\sinh^{2}\left(  \pi qa\right)  }\right] .
\]
The above expression is composed of oscillating terms damped by exponentials.
This is better seen taking the limit for $qa\gg1$:%
\begin{equation}
F(q)\simeq-\frac{4\pi^{2}\rho_{0}a}{q}\left[  R\cos(qR)-\pi
a\sin(qR)\right]  \ e^{-\pi qa}. \label{distsym}
\end{equation}

\begin{figure}
[ptb]
\begin{center}
\includegraphics[
height=2.8461in,
width=3.3364in
]%
{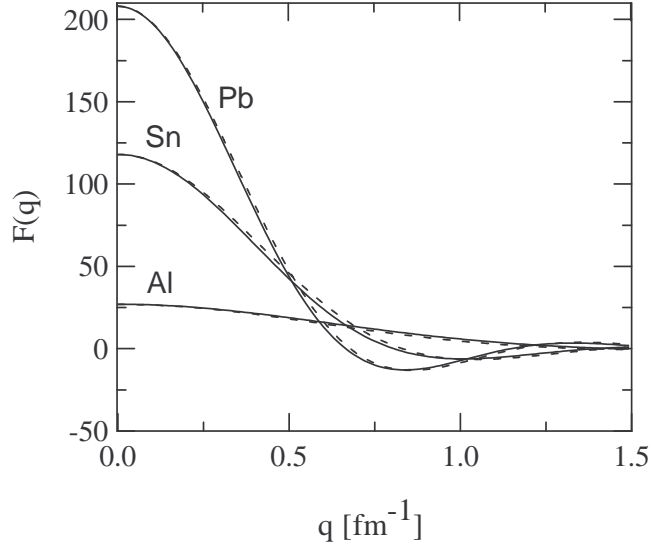}%
\caption{Comparison between form factors of a convoluted Fermi-Yukawa
distribution calculated numerically using eq. \ref{formff} (solid lines), or
with the analytical equation \ref{F0} (dashed lines).}%
\label{fig:form}%
\end{center}
\end{figure}

The Fermi distribution
\begin{equation}
\rho(r)=\rho_{0}/\left[  1+\exp\left\{  \left(
r-R\right)  /a\right\}  \right] \label{formff}
\end{equation}
with central density $\rho_{0}$,
radius $R,$ and diffusiveness $a,$ gives a good description of the
densities of heavy nuclei. This distribution can be fitted by
the convolution $\rho\left(  \mathbf{r}\right)  =\int
d^{3}r^{\prime}\rho_{p}\left(  \mathbf{r}\right)  \rho_{A}\left(
\mathbf{r-r}^{\prime}\right)  $ of a hard sphere for $\rho_{A}\left(
\mathbf{r-r}^{\prime}\right)  $ and an Yukawa function \cite{DN76} for
$\rho_{p}\left(  \mathbf{r}\right)  $. The advantage is that the form factor
factorizes, $F(q)=F_{p}(q)F_{A}(q)$, and
can be calculated analytically as
\begin{equation}
F(q)=\frac{4\pi\rho_{0}}{q^{3}}\left[  \sin\left(  qR\right)  -qR%
\cos\left(  qR\right)  \right]  \left[  \frac{1}{1+q^{2}a_{Y}^{2}}\right]
. \label{F0}%
\end{equation}
The term inside the first parenthesis comes from the hard sphere
(uniform distribution with sharp cutoff at $R$) and the second
parenthesis is the damping term due to an Yukawa diffuseness surface
with width $a_{Y}$.
Figure \ref{fig:form} compares $F(q)$ of eq. \ref{F0} with that obtained with the
numerical integration  of a two-point Fermi density distribution, eq.
\ref{formff}. The
parameters for Al ($R=3.07$ fm , $a=0.519$ fm), Cu ($R=4.163$ fm ,
$a=0.606$ fm), Sn ($R=5.412$ fm , $a=0.560$ fm), Au ($R=$ 6.43 fm ,
$a=0.$ 541 fm) , and Pb ($R=6.62$ fm , $a=0.546$ fm) (to simplify the
figure, I did not plot the curves for Au and Cu). In all cases I used for
the Yukawa function parameter in eq. (\ref{F0}) $a_{Y}=0.7$ fm. We see that
the agreement is very good.

For light nuclei, it is more appropriated to use Gaussian densities,
with the Gaussian parameter $a=\sqrt{\frac{2}{3}\left\langle
r^{2}\right\rangle _{ch}}$, where $\sqrt{\left\langle
r^{2}\right\rangle _{ch}} $ is the root-mean-square radius of the
matter density. \ For example, for carbon, $a=2.018$ fm. For a
Gaussian density parametrized as $\rho\left(  r\right)
=\rho_{0}\exp\left(  -r^{2}/a^{2}\right)  $, one gets%
\begin{equation}
F(q)=\left(  \pi a^{2}\right)  ^{3/2}\rho_{0}\exp\left(  -q^{2}a^{2}/4\right)
\;. \label{F02}%
\end{equation}
To simulate nodes in the wavefunctions of light nuclei, a sum of gaussian
distributions can be used, including terms proportional to $r^{n}\exp\left(
-r^{2}/a^{2}\right)  $. The form factors arising from these terms can be
obtained from nth-order derivatives of eq. \ref{F02} with respect to $1/a^{2}
$.

For halo nuclei the following parametrization can be adopted%
\begin{equation}
\rho\left(  r\right)  =\rho_{1}\exp\left(  -r^{2}/a_{1}^{2}\right)  +\rho
_{2}\left(  \frac{a_{2}}{r}\right)  \exp\left(  -r/a_{2}\right)  ,
\label{denshalo}%
\end{equation}
where the first term describes the density of the core and the
second term describes the extended halo density. The second term blows up as
$r \rightarrow 0$ and has only a meaning in the description of the long
tail characterizing the halo wavefunction. If the core
contains $A_{1}$ nucleons and the halo contains $A_{2}$ nucleons,
the form factor for the above
distribution becomes%
\begin{equation}
F(q)=A_{1}\exp\left(  -q^{2}a_{1}^{2}/4\right)  +\frac{A_{2}}{1+a_{2}^{2}%
q^{2}}. \label{formhalo}%
\end{equation}

\acknowledgments{The author is grateful to Haik Simon for beneficial discussions.
This work was supported by the U.\thinspace S.\
Department of Energy under grant No. DE-FG02-04ER41338.}

\end{document}